\documentclass[fleqn,usenatbib]{mnras}

\usepackage{newtxtext,newtxmath}

\usepackage[T1]{fontenc}

\DeclareRobustCommand{\VAN}[3]{#2}
\let\VANthebibliography\thebibliography
\def\thebibliography{\DeclareRobustCommand{\VAN}[3]{##3}\VANthebibliography}

\usepackage{graphicx}	
\usepackage{amsmath}	

\RequirePackage{siunitx}
\RequirePackage{xspace}

\DeclareSIUnit \century{century}
\DeclareSIUnit \parsec{pc}
\DeclareSIUnit \year{yr}
\DeclareSIUnit \steradian{sr}
\DeclareSIUnit \solarmass{\text{\ensuremath{{M}_{\odot}}}}
\DeclareSIUnit \solarluminosity{\text{\ensuremath{{L}_{\odot}}}}
\DeclareSIUnit \h{\ensuremath{\mathnormal{h}}}
\DeclareSIUnit \hubble{\h}
\DeclareSIUnit \hseventy{\ensuremath{\mathnormal{h_{70}}}}
\DeclareSIUnit \hMpc{\per\h\mega\parsec}
\DeclareSIUnit \micron{\micro\meter}
\DeclareSIUnit \kms{\kilo\meter\per\second}
\DeclareSIUnit \kmsMpc{\kilo\meter\per\second\per\mega\parsec}

\newcommand*{\degree}{\si{\degree}}
\newcommand*{\arcminute}{\si{\arcminute}}
\newcommand*{\arcsecond}{\si{\arcsecond}}

\newcommand*{\hMpc}{\si{\hMpc}}
\newcommand*{\kms}{\si{\kilo\meter\per\second}}
\newcommand*{\kmsMpc}{\si{\kmsMpc}}

\newcommand*{\ra}[2][]{{
    \DeclareSIUnit \degree{\textsuperscript{h}}%
    \DeclareSIUnit \arcminute{\textsuperscript{m}}%
    \DeclareSIUnit \arcsecond{\textsuperscript{s}}%
    \ang[#1]{#2}}}%

\newcommand*{\Euclid}{\textit{Euclid}\xspace}

\newcommand{\sfont}[1]{{\scriptscriptstyle\rm #1}}

\newcommand{\IE}{\ensuremath{I_\sfont{E}}\xspace}
\newcommand{\YE}{\ensuremath{Y_\sfont{E}}\xspace}
\newcommand{\JE}{\ensuremath{J_\sfont{E}}\xspace}
\newcommand{\HE}{\ensuremath{H_\sfont{E}}\xspace}
\newcommand{\BGE}{\ensuremath{{\rm BG}_\sfont{E}}\xspace}
\newcommand{\RGE}{\ensuremath{{\rm RG}_\sfont{E}}\xspace}

\newcommand*{\msun}{\si{\solarmass}}
\newcommand*{\lsun}{\si{\solarluminosity{}}}

\newcommand{\quasar}{EUCL\,J181530.01$+$652054.0}
\newcommand{\quasarshort}{EUCL~QSO\,J1815$+$6520}
\newcommand{\noquasar}{EUCL\,J180409.14$+$641335.3}

\newcommand{\mdwarf}{EUCL\,J174429.80$+$672728.1}
\newcommand{\dwarfabell}{EUCL\,J002516.31$-$491618.5}

\def\siiv     {\ensuremath{\text{Si\,\textsc{iv}}}}

\def\civ     {\ensuremath{\text{C\,\textsc{iv}}}}

\def\mgii     {\ensuremath{\text{Mg\,\textsc{ii}}}}

\def\ciii     {\ensuremath{\text{C\,\textsc{iii]}}}}

\def\halpha {\ensuremath{\text{H}\alpha}}
\def\hbeta {\ensuremath{\text{H}\beta}}

\def\nv     {\ensuremath{\text{N\,\textsc{v}}}}

\newcommand*{\AckInstitutions}{a number of agencies and
  institutes that have supported the development of \Euclid, in
  particular
  the Agenzia Spaziale Italiana,
  the Austrian Forschungsf\"orderungsgesellschaft funded through BMK,
  the Belgian Science Policy,
  the Canadian Euclid Consortium,
  the Deutsches Zentrum f\"ur Luft- und Raumfahrt,
  the DTU Space and the Niels Bohr Institute in Denmark,
  the French Centre National d'Etudes Spatiales,
  the Funda\c{c}\~{a}o para a Ci\^{e}ncia e a Tecnologia,
  the Hungarian Academy of Sciences,
  the Ministerio de Ciencia, Innovaci\'{o}n y Universidades,
  the National Aeronautics and Space Administration,
  the National Astronomical Observatory of Japan,
  the Netherlandse Onderzoekschool Voor Astronomie,
  the Norwegian Space Agency,
  the Research Council of Finland,
  the Romanian Space Agency,
  the State Secretariat for Education, Research, and Innovation (SERI) at the Swiss
  Space Office (SSO),
  and the United Kingdom Space Agency.
  A complete and detailed list is available on the \Euclid\ web site
  (\url{www.euclid-ec.org}).\xspace}

  \newcommand{\AckEC}{The Euclid Consortium acknowledges the European
  Space Agency and \AckInstitutions}

\title[A $z=5.4$ quasar found by \Euclid\ spectroscopy]{\Euclid: The potential of slitless infrared spectroscopy: \\A $z=5.4$ quasar and new ultracool dwarfs\thanks{This paper is published on behalf of the Euclid Consortium}}

\author[E.~Ba\~nados et al.]{\footnotesize
E.~Ba\~nados$^{1}$\thanks{E-mail: banados@mpia.de},
V.~Le~Brun$^{2}$\thanks{VLB would like to dedicate this paper to the memory of Bianca Garalli, who passed away on September 20, 2024, for her invaluable contribution to the \Euclid\ spectroscopic pipeline.},
S.~Belladitta$^{1,3}$,
I.~Momcheva$^{1}$,
D.~Stern$^{4}$,
J.~Wolf$^{1}$,
M.~Ezziati$^{2}$,
D.~J.~Mortlock$^{5,6}$,
A.~Humphrey$^{7,8}$,
R.~L.~Smart$^{9,10}$
\newauthor \footnotesize
S.~L.~Casewell$^{11}$,
A.~P\'erez-Garrido$^{12}$
% ...
 \\
 \\
\emph{Affiliations can be found at the end of the article.}\\
\vspace{-1.5cm}
}

\pubyear{\the\year{}}

\begin{document}
\label{firstpage}
\pagerange{\pageref{firstpage}--\pageref{lastpage}}
\maketitle


\begin{abstract}
\footnotesize
We demonstrate the potential of \Euclid{\rm 's} slitless spectroscopy to discover high-redshift ($z>5$) quasars and their main photometric contaminant, ultracool dwarfs. Sensitive infrared spectroscopy from space is able to efficiently identify both populations, as demonstrated by \Euclid\ Near-Infrared Spectrometer and Photometer Red Grism (NISP \RGE) spectra of the newly discovered $z=5.404$ quasar \quasar, as well as several ultracool dwarfs in the Euclid Deep Field North and the \textit{Euclid} Early Release Observation field Abell~2764. 
The ultracool dwarfs were identified by cross-correlating their spectra with templates. The quasar was identified by its strong and broad \ciii\ and \mgii\ emission lines in the NISP \RGE\ 1206--1892\,nm spectrum, and confirmed through optical spectroscopy from the Large Binocular Telescope. The NISP Blue Grism (NISP \BGE) 926--1366\,nm spectrum confirms \civ\ and \ciii\ emission. 
NISP \RGE\ can find bright quasars at $ z\approx 5.5$ and $ z\gtrsim 7$, redshift ranges that are challenging for photometric selection due to contamination from ultracool dwarfs. 
\quasar\ is a high-excitation, broad absorption line quasar detected at 144\,MHz by the LOw-Frequency Array ($L_{\rm 144}=4.0 \times 10^{25}\,$W\,Hz$^{-1}$). The quasar has a bolometric luminosity of $3\times 10^{12}\,\lsun$ and is powered by a $3.4\times 10^9\,\msun$ black hole. The discovery of this bright quasar is noteworthy as fewer than one such object was expected in the $\approx$20\,deg$^2$ surveyed. This finding highlights the potential and effectiveness of NISP spectroscopy in identifying rare, luminous high-redshift quasars, previewing the census of these sources that \Euclid's slitless spectroscopy will deliver over about $14\,000\,$deg$^2$ of the sky. 
\end{abstract}

\clearpage
\begin{keywords}
Galaxies: quasars:individual \quasar\ ---  Stars: brown dwarfs ---  Stars: individual:  \mdwarf, \dwarfabell
\end{keywords}



\section{\label{sc:Intro}Introduction}
Quasars are accreting supermassive black holes in the centres of massive galaxies that can be studied in detail at large cosmological distances, even within the first Gyr after the big bang.  These distant quasars provide important constraints on the formation and growth of supermassive black holes, massive galaxies, the build-up of large-scale structure, and the Universe's last major phase transition, the epoch of reionisation \citep[see][for a recent review]{fan2023}. 

Quasars at $z\gtrsim 5$ have traditionally been identified from photometric colour selections \citep[e.g.,][]{jiang2016,matsuoka2019,belladitta2025} assisted by machine-learning and probabilistic approaches \citep[e.g.,][]{mortlock2012,wenzl2021,byrne2024}. 
Candidates are then confirmed through spectroscopic observations \citep[e.g.,][]{yang_daming2024}. The main challenges to identifying the most distant quasars are  (i) the rapid decline of their number density at $z>5$ \citep[e.g.,][]{schindler2023,matsuoka2023}; and (ii) the similar colours of the more abundant late M and L and T brown dwarf populations. 
Selection effects produce a lack of quasars at $z\approx 5.5$ and between $z=7.1$ \citep{mortlock2011} and $z=7.5$ \citep{banados2018a,yang2020,Wang2021ApJ...907L...1W}. The first gap is due to the colours of $z\approx 5.5$ quasars being almost indistinguishable from some M and L dwarfs, the most abundant stars in our Galaxy (see, e.g., Figs.~1 in \citealt{banados2016} and \citealt{matsuoka2016}). Most of the $z\approx5.5$ quasars known have been discovered through dedicated campaigns to fill this gap \citep[e.g.,][]{yang2019a}. The second gap centred at $z\approx 7.3$ is due to the photometric contamination of L and T dwarfs (see, e.g., \citealt{hewett2006, lodieu2007,mortlock2009,burningham2013} and Fig.~2 in \citealt{fan2023}). Currently, there are more than $11\,000$ spectroscopically confirmed M6--M9, $\sim 2200$ L, and $\sim 800$ T ultracool dwarfs \citep{smart2019,best_2024_10573247}.

\begin{figure*}
    \centering
     \includegraphics[width=0.99\linewidth]{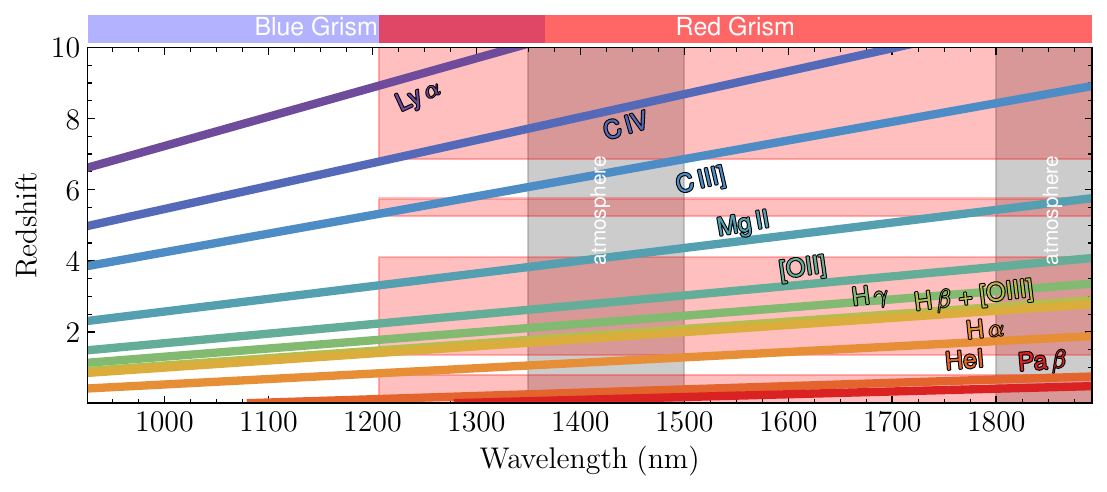}
    \caption{
    Quasar strong emission lines as a function of redshift within the NISP \BGE\ and \RGE\ bandpasses \citep{EuclidSkyNISP}, as indicated at the top of the figure. NISP \BGE\ data will only be available in the EDS, while NISP \RGE\ data will be present in the EWS.  The vertical grey-shaded regions correspond to wavelengths of strong telluric absorption, where ground-based telescopes are not sensitive. The horizontal red-shaded regions represent the redshift ranges with at least two strong emission lines expected in the NISP \RGE\ spectral bandpass, thereby providing the most reliable redshifts for `blind' discoveries over the entire EWS. 
  }
    \label{fig:motivation}
\end{figure*}

The next breakthrough for reionisation-era quasar discoveries is expected to come from the \Euclid\ mission \citep{EuclidSkyOverview}. 
The Euclid Wide Survey (EWS; \citealt{Scaramella-EP1}) will cover about $14\,000$\,deg$^2$ of extragalactic sky in the optical (\IE\ filter; \citealt{EuclidSkyVIS}) and near-infrared (\YE, \JE, and \HE\ filters; \citealt{EuclidSkyNISP,Schirmer-EP18}). The expected quasar yields from \Euclid\ photometric selection are discussed for  $z<7$ in \cite{EP-Selwood} and for $z>7$ by \cite{Barnett-EP5}. Figure~\ref{fig:color-color} shows that photometric contamination of brown dwarfs is also expected to be one of the main challenges for $z>5$ quasar identification using only \Euclid\ photometry. 

In addition to photometry, the Near-Infrared Spectrometer and Photometer (NISP) on \Euclid\ also provides grism slitless spectroscopy with a resolving power greater than 480 (for a $0\farcs5$ diameter object) over the range 1206--$1892\,$nm \citep[referred to as the red grism; \RGE;][]{EuclidSkyNISP}. The \RGE\ data is available throughout the entire EWS, while NISP also offers blue grism spectroscopy (\BGE) with a resolving power greater than 400 (for a $0\farcs5$ diameter object) over the range 926--$1366\,$nm, exclusively in the Euclid Deep Survey (EDS), covering approximately 60\,deg$^2$ \citep{EuclidSkyNISP, EuclidSkyOverview}.  
Here, we discuss and demonstrate the potential for discovering quasars (and their contaminants) directly from NISP spectroscopy. 
{Given the low number density of bright quasars at $z>5$, with only a few expected per 100\,deg$^2$ (\citealt{schindler2023,matsuoka2023}), we primarily focus on the capabilities of the NISP \RGE, which spans the largest area where significant discoveries are anticipated. 
Figure~\ref{fig:motivation} shows the strongest quasar emission lines that fall within the NISP grism wavelength range as a function of redshift. When more than one strong emission line is observed, the quasar nature and redshift can, in principle, be obtained directly from the NISP spectrum.

\begin{figure}
    \centering
    \includegraphics[width=0.975\linewidth]{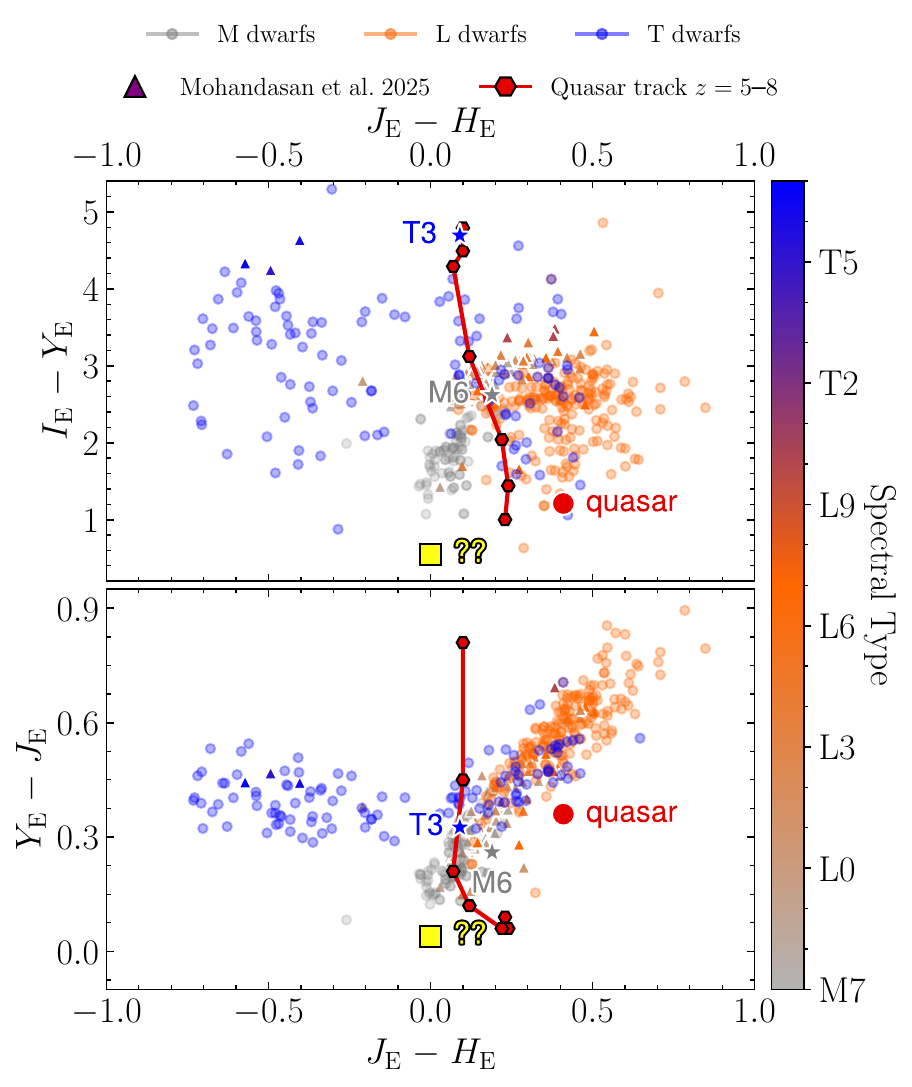}
    \caption{
      Top: (Bottom:) $\IE - \YE$ ($\YE - \JE$) versus \ $\JE - \HE$ 
     diagram showing the synthetic colours of the M-, L-, and T-dwarfs from the SpeX Prism Library (small circles), 
     as measured by \protect\cite{EROLensVISDropouts}.
     The triangles represent 
     ultracool dwarfs discussed in \protect\cite{Q1-SP042}, with 
     spectral types indicated by the colour bar. The solid line represents the colour track of the $z=6$ 
     quasar composite spectrum of \protect\cite{banados2016} 
     combined at rest-frame 1300\,\AA\ with the average spectrum of \protect\cite{vandenberk2001}. 
     The hexagonal markers are plotted in steps of $\Delta\, z =0.5$, starting at $z=5$ at the bottom and finishing at $z=8$ at the top. The larger, labelled symbols represent the colours of the individual sources discussed in this paper: the T3 dwarf \dwarfabell\ 
     (blue star), the M6 dwarf 
     \mdwarf\ (grey star), the $z=5.4$ quasar
     \quasar\ (red circle), and the unidentified source 
     \noquasar\ (yellow square). 
     These colour-colour diagrams are for context, and we emphasise that \textit{Euclid} photometry was not used to identify these sources (except the T3 dwarf; see Sect.~\ref{sec:a2764}). 
    }
    \label{fig:color-color}
\end{figure}

This paper is structured as follows. 
We describe the \Euclid\ data used for this work in Sect.\,\ref{sec:grism}. 
In Sect.\,\ref{sec:discovery-dwarfs}, we briefly discuss the discovery potential for ultracool dwarfs.  
In Sect.\,\ref{sec:discovery-quasars}, we introduce two $z\approx 5.5$ quasar candidates identified with NISP \RGE.  We discuss the properties of a newly discovered $z=5.4$ quasar in Sect.\,\ref{sec:quasar}.  
Finally, in Sect.\,\ref{sec:outlook} we provide a summary and highlight additional science cases enabled by NISP slitless spectra. 
Appendix\,\ref{apdx:A} lists the ultracool dwarf templates used in this work. Appendix\,\ref{sec:2dspec} provides an example of NISP \RGE\ two-dimensional spectrograms. In Appendix\,\ref{sec:bluegrism}, we show NISP \BGE\ spectra of the $z\approx 5.5$ quasar candidates. 
We adopt a standard, flat cosmological model with $H_0 = 70 \,\mbox{km\,s}^{-1}$\,Mpc$^{-1}$ and \mbox{$\Omega_{\rm m} = 0.30$}. All \Euclid\ magnitudes reported are from aperture photometry in the AB system unless otherwise stated. All postage stamps are oriented north up and east to the left.

\section{Data}\label{sec:grism}

In this paper, we use NISP \RGE\ data from the phase verification campaign in the Euclid Deep Field North (EDF-N; 20\,deg$^2$ centred on $\rm{RA}=\ra{17;58;55.9}$ and $\rm{Dec}= \ang{+66;01;04.7}$) and from the Early Release Observations \citep[ERO;][]{EROcite} of the lensing cluster Abell 2764, centred on $\rm{RA}=\ra{00;22;50.1}$ and $\rm{Dec}= \ang{-49;15;59.8}$ \citep{EROLensData}.  

The EDF-N NISP \RGE\ grism data were the first validated and made available to the Euclid Consortium. Near the completion of this work, the EDF-N NISP \BGE\ grism data were made available to the Euclid Consortium for validation. In Appendix \ref{sec:bluegrism}, we showcase some of the first NISP \BGE\ spectra, and we note that these spectra are not available in the first \Euclid Quick Data Release \citep[Q1;][]{Q1-TP001}. The data used here are from one Reference Observing Sequence (ROS), equivalent to the depth expected for the EWS. The NISP grism data have been fully processed with the standard \Euclid\ pipeline \citep[see Sect.~7.5 in][]{EuclidSkyOverview}. 
We use the merged 
catalogue from the phase verification campaign (\texttt{mer-pv}) for coordinates, photometry, and $\texttt{OBJECT-ID}$ \citep[for details, see Sect.~7.4 in][]{EuclidSkyOverview}. 
The EDF-N also has dedicated deep radio 144\,MHz observations (with central RMS noise of 32\,$\mu$Jy\,beam$^{-1}$) from the LOw-Frequency Array (LOFAR; \citealt{bondi2024}). 

The grism data in the Abell 2764 field are from three ROS. However, the ERO data have not been processed through the standard \Euclid\ pipeline. Indeed, only imaging data products have been published so far, reduced with a custom-made pipeline (\citealt{EROData}). All magnitudes reported in this field are from the catalogue presented in \cite{EROLensVISDropouts}. To extract spectra, we performed the following steps. We mosaiced the 16 individual detectors, both for the direct and for the dispersed images, into single images using the world coordinate system from the \HE-band exposures. We derived ad-hoc trace equations for all four grism settings using bright stars, mapping the ($x_{\rm g}$, $y_{\rm g}$) positions of the start of the spectra and the spectral slope as a function of the ($x_{\rm i}$, $y_{\rm i}$) positions of the objects in the direct image. Two-dimensional cut-outs were extracted for each object. Spectra with the same grism-angle combinations from different ROS were rectified, combined together and background-subtracted. One-dimensional spectra were extracted using a box-car extraction aperture of seven pixels. 
The wavelength was calibrated against a handful of emission line objects with known redshifts in the Abell 2764 field. Even though the variation of the wavelength solution across the field has not been mapped in detail, the above model proves sufficiently accurate for the current investigation  (see Sect.~\ref{sec:a2764}).

\section{Identification of ultracool dwarfs with \Euclid\ NISP grism}\label{sec:discovery-dwarfs}

Our primary goal is to use NISP \RGE\ data to identify quasars at $z\approx5.5$ and $z\gtrsim 7$ (Fig.~\ref{fig:motivation}). 
However, current estimates of the quasar luminosity function \citep{schindler2023,matsuoka2023} suggest that there are fewer than 0.3 and 0.04 bright quasars (with $M_{1450}<-25.5$)  in about $20\,$deg$^2$ at $z\approx5.5$  and $z\gtrsim 7$, respectively. 
Consequently, the initial goal of this study was to examine the NISP spectra of typical contaminants for high-redshift quasars, particularly the far more numerous ultracool dwarfs.

Thus, in addition to the templates used to determine redshifts through template fitting described in Sect.~7.5.2 of \cite{EuclidSkyOverview}, we include M-, L-, and T-dwarf templates from the SpeX Prism Spectral Libraries \citep{burgasser2014}, which we list in Appendix~\ref{apdx:A}. The main limitation of the present work is the restricted number of templates used for classification. However, resampling the spectra, we find that our classification is robust within $\pm 1$ spectral type. 
Indeed, this experiment recovered known ultracool dwarfs in the field and enabled the discovery and confirmation of 33 new ones ranging from M7 to T1, which are presented in detail in \cite{Q1-SP042}.  In this paper, we will present examples of two new ultracool dwarfs, which are not in the sample of \cite{Q1-SP042}.

\subsection{A new T3 dwarf in the ERO field Abell 2764}
\label{sec:a2764}
We selected \dwarfabell\ as a potential high-redshift quasar candidate based on a large $\IE-\YE>4$ colour and flat NISP colours (see Fig.~\ref{fig:color-color} and top panel of Fig.~\ref{fig:a2769}). We used the photometry reported in the catalogue of \cite{EROLensVISDropouts} ($\texttt{CATALOG ID}=373511$).

The source \dwarfabell\ is the brightest and one of the most promising $z>6$ quasar candidates in the Abell~2764 field. However, the chances of identifying a $\YE<19$ quasar at $z>6$ in just 0.75\,deg$^2$ are negligible \citep{schindler2023,matsuoka2023}. If this were indeed a quasar, it would be among the most luminous sources ever reported in the early Universe \citep{wu2015,fan2019}. 
To confirm or refute this potentially remarkable serendipitous discovery, we developed our own pipeline to extract the \Euclid\ NISP spectrum of these ERO data (see Sect.~\ref{sec:grism}).  
Figure~\ref{fig:a2769} shows the extracted spectrum, which clearly classifies the source as a brown dwarf.  
We note that this object was photometrically identified as a T3 candidate by \cite{delponte2023}.  
Resampling the spectrum reveals that the best match template varies between T3 and T4, although visually, neither template is a perfect match. The template of a T3 binary, 2MASSJ12095613$-$1004008 \citep{burgasser2004,dupuy2012} appears to be a visually better match (plotted in Fig.~\ref{fig:a2769}), suggesting that it could also be a T binary. 

\begin{figure*}
    \centering
    \includegraphics[width=0.8\linewidth]{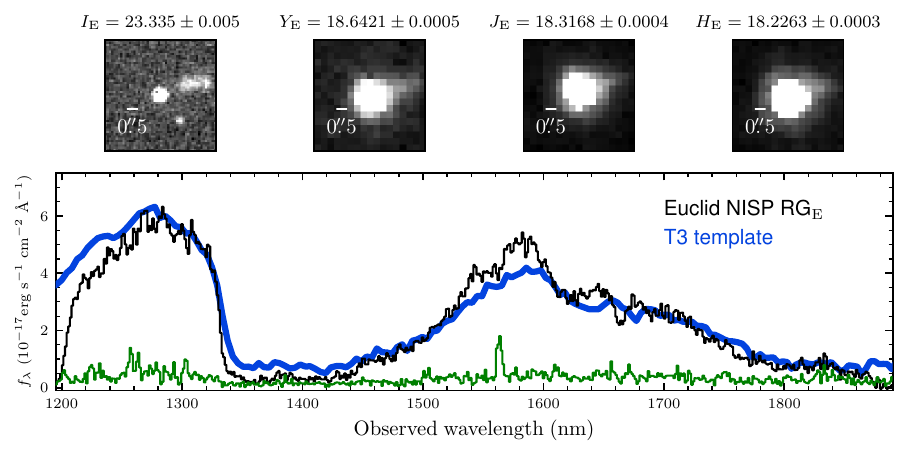}
    \caption{
    \textit{Top:} Postage stamps of the T3 dwarf  \dwarfabell.  The \Euclid\ \IE, \YE, \JE, and \HE\ images are $5\arcsec$ on a side.    
    \textit{Bottom:} NISP \RGE\ grism spectrum (black line and the uncertainties in green). The blue line shows a template of a T3 binary: 2MASS\,J12095613$-$1004008 \citep{burgasser2004,dupuy2012}. 
    }
    \label{fig:a2769}
\end{figure*}

\subsection{A new M6 dwarf in the EDF-N}
\label{sec:lbt-dwarf}

Object \mdwarf\ ($\texttt{OBJECT-ID}=2661241859674578148$) was first selected as a $z\approx 5.6$ quasar candidate from the Pan-STARRS1 survey \citep{banados2023}, but rejected as a quasar after a follow-up observation with the Multi-Object Double Spectrograph (MODS; \citealt{pogge2010}) at the Large Binocular Telescope (LBT). The LBT/MODS observations were carried out in dual mode on 2017 April 21 and June~5. The red grating G670L and a 1\farcs2 slit were used for a total exposure time of one hour.   We present the LBT spectrum for the first time here (the bottom panel of Fig.~\ref{fig:mdwarf}).

The MODS spectrum was reduced with the open-source \textsc{python}-based Spectroscopic Data Reduction Pipeline \textsc{pypeit}\footnote{\url{https://github.com/pypeit/PypeIt}} \citep[version 1.14.1;][]{Prochaska2020}. 
With that pipeline, we perform image processing, including gain correction, bias subtraction, and flat fielding.  
The extracted spectrum was flux-calibrated with a sensitivity function derived from the observation of a spectroscopic standard star. 
The spectra were then co-added and absolute flux calibrated to match the \IE\ magnitude.

Since this quasar candidate is located in the EDF-N, we analysed the \Euclid\ grism spectrum independently of the existing LBT spectrum. The best-fitting template was an M6 dwarf, shown in the middle panel of Fig.~\ref{fig:mdwarf}. 
The M6 dwarf template is relatively featureless in the 1200--$1900\,$nm regime. Notably, the same template reproduces the optical features seen in the LBT spectrum (bottom panel of Fig.~\ref{fig:mdwarf}). 
By resampling the \Euclid\ spectrum, we find that approximately 95\% of the cases classify this source as an M6, 4\% as an M7, and 1\% as an M8 type. 
 \Euclid\ cutouts and photometry are shown in the top panel of Fig.~\ref{fig:mdwarf}.

\begin{figure*}
    \centering
    \includegraphics[width=0.8\linewidth]{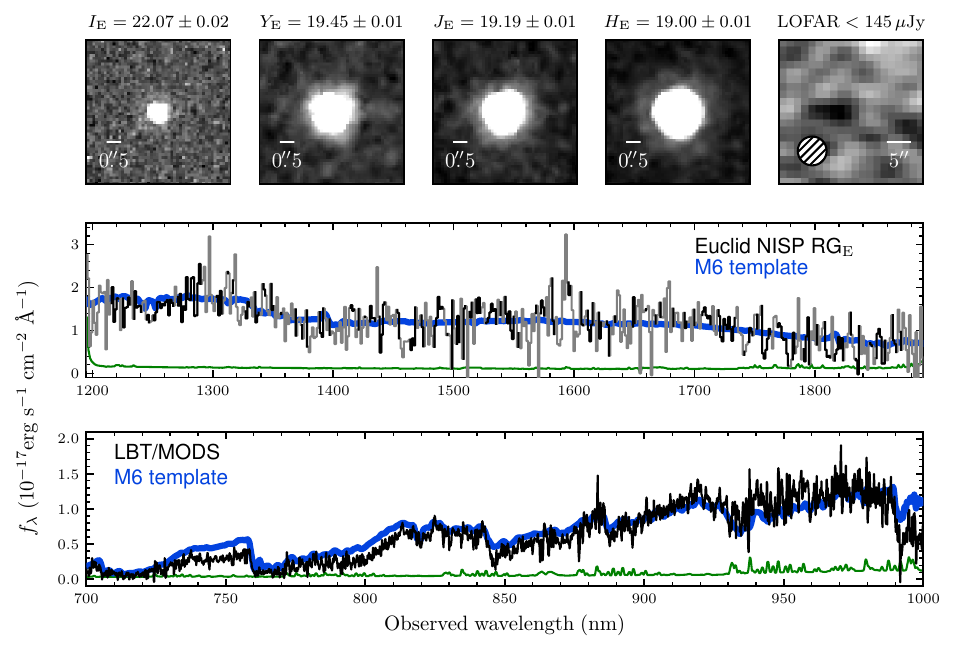}
    \caption{
    \textit{Top:} Postage stamps of the M6 dwarf  \mdwarf. The \Euclid\ \IE, \YE, \JE, and \HE\ images are $5\arcsec$ on a side, while the LOFAR image is $30\arcsec$ on a side. The LOFAR   beam is shown in the lower left of its panel and the reported flux density corresponds to a $3\,\sigma$ upper limit.  
    \textit{Middle:}  NISP \RGE\ grism spectrum (black line; masked pixels are in grey, and the uncertainties in green). The blue line shows the best template fit, identifying this as an M6 dwarf. 
    \textit{Bottom: } LBT/MODS optical spectrum (black line and uncertainties in green), confirming the \Euclid\ classification. 
    The spectral features are clearly well-matched to the observed optical spectrum. 
    The template corresponds to LHS~36 (also known as Wolf~359), originally published in \protect\cite{burgasser2008}.
    }
    \label{fig:mdwarf}
\end{figure*}

\section{Identification and follow-up observations of $z>5$ \Euclid\ quasar candidates}\label{sec:discovery-quasars}

We added the quasar composite spectrum of \cite{vandenberk2001} as part of the redshift template fitting of the \Euclid\ pipeline in order to be able to identify $z>5$ quasars. In the EDF-N field, there were only two sources for which the quasar template was the best match and at an implied redshift where two strong emission lines are expected in the NISP \RGE\ spectra (Fig.~\ref{fig:motivation}). This was a selection based purely on the NISP \RGE\ spectra matched to templates; no photometry or additional information was used. For completeness, we show the NISP \BGE\ spectra of these sources in Appendix \ref{sec:bluegrism}.

\subsection{\quasar}

The quasar template proved to be the best match to the source 
\quasar\ (hereafter \quasarshort; $\texttt{OBJECT-ID}=2738750478653483354$), 
implying a quasar at $z=5.40$. 
The \Euclid\ spectra, photometry, and cutout images of \quasarshort\ are displayed in the top panel of Fig.~\ref{fig:qso}. We visually inspected the \Euclid\ spectrum and found that the \ciii\ and \mgii\ lines were robustly detected and could be well-fitted by single Gaussians (Fig.~\ref{fig:qso}), with a \mgii-redshift of $z_{\rm \mgii}=5.404\pm 0.007$ (the age of the Universe was 1.04\,Gyr at this redshift).

We observed \quasarshort\ with LBT/MODS on 2024 June 17. The observations were carried out in dual mode with the red grating G670L, a slit of 1\arcsec\ width, and a total exposure time of 15 min. The spectrum was reduced as described in Sect.~\ref{sec:lbt-dwarf} and is shown in the bottom panel of Fig.~\ref{fig:qso}, confirming the quasar nature of \quasarshort.  
The spectrum reveals a \civ\ line with an equivalent width of 
$(13\pm0.4)\,$\AA\ and blueshifted by 
$(2070\pm330)\,$\kms\ 
with respect to the \mgii\ line, 
consistent with quasars displaying strong broad line region outflows \cite[e.g.,][]{vietri2018,rankine2020,gillette2024}. 
We measured the rest-frame absolute magnitude at 1450\,\AA\ directly from the LBT spectrum, resulting in $M_{1450}=-25.52 \pm 0.01$. 

\begin{figure*}
    \centering
    \includegraphics[width=0.8\linewidth]{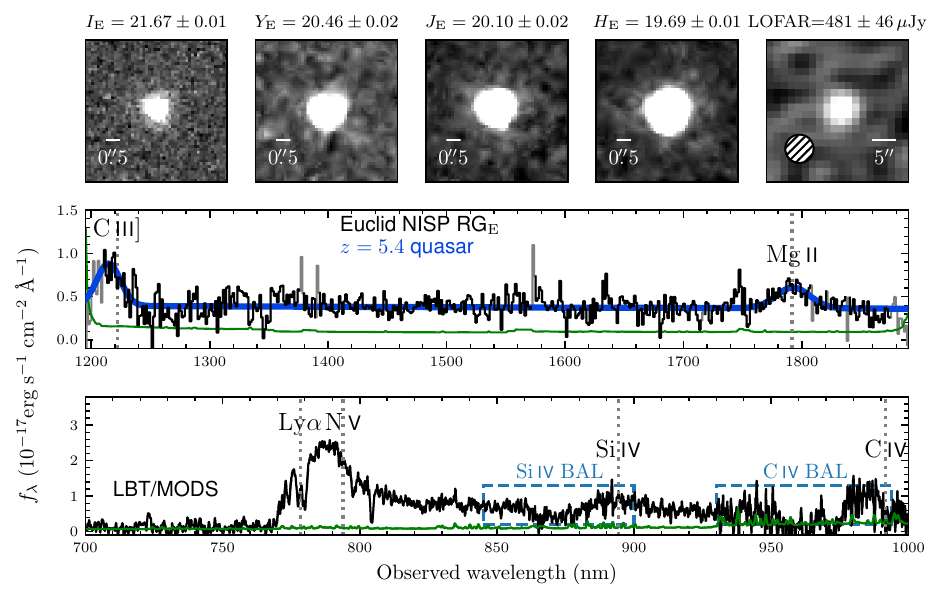}
    \caption{
    \textit{Top:} Postage stamps of the $z=5.4$ quasar \quasar. The \Euclid\ \IE, \YE, \JE, and \HE\ images are $5\arcsec$ on a side while the LOFAR image is $30\arcsec$ on a side. The LOFAR   beam is shown on the lower left of its panel.  
    \textit{Middle:}  NISP grism \RGE\ spectrum (black line; masked pixels are in grey, and the uncertainties in green). The blue line shows the best-fitting power-law emission plus \ciii\ and \mgii\ broad emission lines, identifying this as a quasar at $z_{\rm \mgii}=5.404\pm 0.007$. 
    \textit{Bottom: } LBT/MODS optical spectrum (black line and uncertainties in green), confirming the quasar nature of \quasar. The vertical dashed lines correspond to the expected position of the labelled emission lines based on the \mgii\ redshift. 
    The dashed rectangles indicate the \siiv\ and \civ\ BAL regions shown in Fig.~\ref{fig:bal}. 
    }
    \label{fig:qso}
\end{figure*}

\subsection{\noquasar}
\label{sec:noquasar}
The template fitting of \noquasar\ ($\texttt{OBJECT-ID}=2710381121642264965$), implied a quasar at $z=5.37$. 
The \Euclid\ spectrum, cutouts, and photometry are shown in the top panel of Fig.~\ref{fig:qso-contaminant}. 

The visual inspection of the spectrum is not as convincing as that of \quasarshort.  The feature that is expected to be \mgii\ at $z=5.37$ is broader than the quasar template and the existence of \ciii\ is unclear (Fig.~\ref{fig:qso-contaminant}).

 To come full circle on testing this quasar-discovery strategy, we obtained follow-up optical spectroscopy with the Double Spectrograph (DBSP; \citealt{oke1982}) on the 5-m Hale telescope at Palomar Observatory on 2024 July 10. We obtained three exposures of $1200\,$s each using the $1\farcs5$ slit. 
The data were reduced analogously to the LBT spectrum described in Sect.~\ref{sec:lbt-dwarf}, but with the \textsc{pypeit} version 1.16.0. The Palomar/DBSP spectrum (the bottom panel of Fig.~\ref{fig:qso-contaminant}) does not show the sharp break expected at $0.77\,\micron$ for a $z=5.4$ quasar (compare with the LBT spectrum in Fig.~\ref{fig:qso}). Indeed, the Palomar spectrum does not reveal any strong emission lines, and is relatively featureless. This spectrum is not well reproduced by any of the current templates used in the \Euclid\ pipeline, and finding the exact spectral classification is beyond the scope of this work. 

\begin{figure*}
    \centering
    \includegraphics[width=0.8\linewidth]{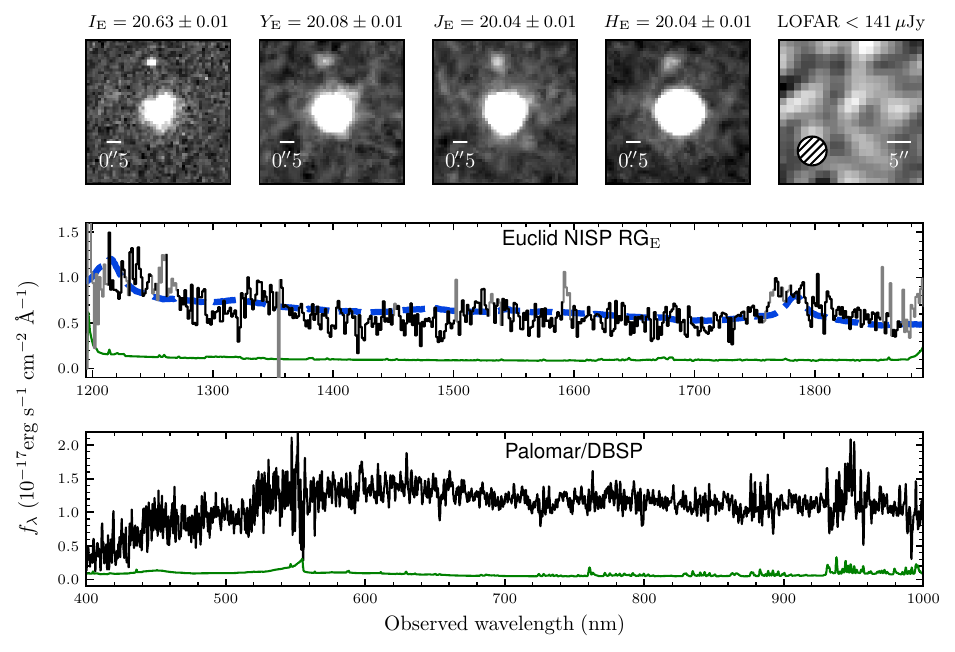}
    \caption{
    \textit{Top:} Postage stamps of the quasar candidate \noquasar. The \Euclid\ \IE, \YE, \JE, and \HE\ images are $5\arcsec$ on a side while the LOFAR image is $30\arcsec$ on a side. The LOFAR   beam is shown in the lower left of its panel and the reported flux density corresponds to a $3\,\sigma$ upper limit. 
    \textit{Middle:} NISP \RGE\ grism spectrum (black line; masked pixels are in grey and the uncertainties in green). The dashed blue line shows the best-fitting template corresponding to a quasar \citep{vandenberk2001} redshifted to $z=5.37$.  
    \textit{Bottom: } Palomar/DBSP optical spectrum (black line and uncertainties in green, revealing a relatively flat spectrum, ruling out \noquasar\ being a $z\sim 5.4$ quasar. 
    }
    \label{fig:qso-contaminant}
\end{figure*}

\section{Physical properties of the  $z=5.4$ \Euclid\ quasar}
\label{sec:quasar}

\subsection{Black hole mass}

The \Euclid\ spectrum of \quasarshort\ covers the broad \mgii\ emission line (middle panel of Fig.~\ref{fig:qso}), which is one of the most reliable tracers to derive single-epoch, black hole mass measurements \citep[e.g.,][]{fan2023}. It is not possible to study \mgii\ from the ground at $z\approx 5.4$ due to the low atmospheric transparency at around $1800\,$nm (Fig.~\ref{fig:motivation}). 

We use the relationship presented in \cite{vestergaard2009} to estimate the black hole mass from the full-width at half maximum of the \mgii\ line [FWHM$_{\rm \mgii}=(5138 \pm  616)\,\kms$] and the luminosity at $3000\,\AA$ [$L_{3000}= (5.7 \pm 0.2)\,\times 10^{12}\,\lsun$], yielding $M_{\rm BH}= (3.0\pm 0.7) \times 10^9\,\msun$.  
Adopting a widely used bolometric correction ($L_{\rm Bol}= 5.15\,L_{3000}$; see e.g., \citealt{mazzucchelli2023}), we find an Eddington ratio of $L_{\rm Bol}\,/\,L_{\rm Edd} = 0.3\pm 0.1$. 
These properties are consistent with the bulk of the quasars studied at $z\gtrsim 5$ \citep[e.g.,][]{shen2019,lai2024}.

\subsection{BAL properties}
\label{sec:bal}

The LBT spectrum of \quasarshort\ (bottom panel of Fig.~\ref{fig:qso}) not only validates the \Euclid\ discovery but also reveals strong absorption features blueward of \nv, \siiv, and \civ, classifying this source as a high-excitation, broad absorption line (BAL) quasar. 
The presence of BALs in quasar spectra indicates strong outflows, launched from accretion discs, that can have velocities of up to 20\% of the speed of light \citep[$c$; e.g.,][]{rodriguez2020}.  
The exact fraction of BAL quasars is still debated, but it ranges from 10 to 50 percent \citep{dai2008,allen2011,bischetti2022}.

The \nv\ BAL in \quasarshort\ coincides with the wavelengths absorbed by foreground neutral hydrogen in the intergalactic medium (Fig.~\ref{fig:qso}). Thus, we cannot determine its velocity structure confidently, and instead we focus on the \siiv\ and \civ\ BALs. 
We use the task {\tt continuumfit} from the \texttt{linetools} python package\footnote{\url{https://github.com/linetools}} to interactively fit the quasar continuum and then normalise its flux. Figure~\ref{fig:bal} shows the normalised spectra around the BAL regions highlighted in Fig.~\ref{fig:qso}.  
The detached and terminal velocities quantify the minimum and maximum outflow velocities of the gas traced by the BAL \citep{hall2002}. 
To be conservative, we measured the BAL minimum detached and terminal velocities from the 90\% level of the normalised spectrum.  We obtained the same range of velocity for both BALs, 0.015--$0.041\,c$, indicating that they originate from the same kinematic region (see Fig.~\ref{fig:bal}). 
This quasar has a \civ\ balnicity index (BI; \citealt{weymann1991}) of $\mathrm{BI}=3766^{+1128}_{-1809}\,\kms$, indicating a powerful outflow \citep[e.g.,][]{bischetti2022}. 

\begin{figure}
    \centering
    \includegraphics[width=0.975\linewidth]{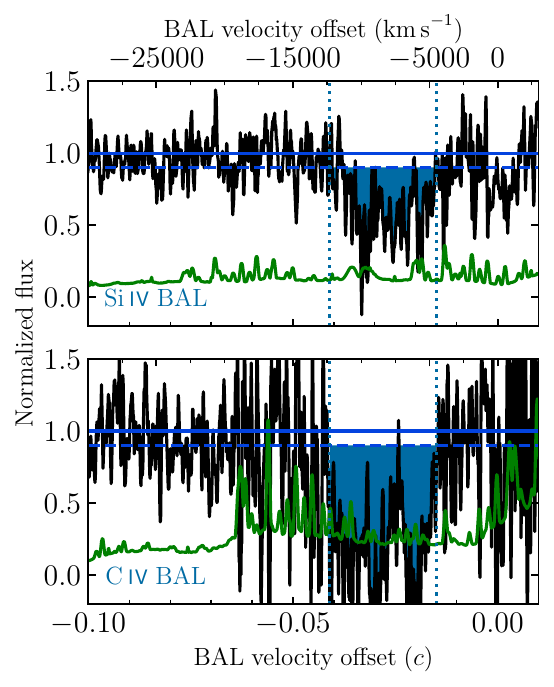}
    \caption{
    Normalised LBT/MODS spectrum of \quasarshort\ zoomed-in on the BAL regions (see Fig.~\ref{fig:qso} and Sect.~\ref{sec:bal}). The solid and dashed horizontal lines correspond to 100\% and 90\% of the normalised spectrum, respectively. The vertical dotted lines and the blue-shaded region show that the \siiv\ (top) and \civ\ (bottom) outflows have velocities $0.015$--$0.041\,c$.
    }
    \label{fig:bal}
\end{figure}

\subsection{Radio properties}
\label{sec:radio}

The quasar \quasarshort\ is well-detected in the LOFAR 144\,MHz data shown in the top panel of Fig.~\ref{fig:qso}. The source is outside of the central circular $10\,$deg$^2$ region used to create the LOFAR-EDF-N catalogue \citep{bondi2024}. Thus, we measured the flux density directly from the beam-corrected image.\footnote{\url{https://lofar-surveys.org/deepfields_public_edfn.html}} The source is unresolved, and we measure a peak flux density of $(481 \pm 46)\,\mu$Jy. We note that the radio data were not used for the selection of the quasar, and that late M-dwarfs can also show comparable radio emission \citep{gloudemans2023}.

Radio-loudness in quasars is an observational parameter used to quantify the power of synchrotron radiation with respect to emission in the UV/optical regime coming from the accretion disc. The radio-loudness is usually defined as the ratio of the flux densities at rest-frame 5\,GHz and 2500\,\AA\ or 4400\,\AA.  
Here we use the former definition, $R_{2500}$, since rest-frame 2500\,\AA\ is covered by the \Euclid\ spectrum, while for 4400\,\AA\ we would need to extrapolate. 
Since we only have a radio detection at 144\,MHz, we extrapolate to rest-frame 5\,GHz assuming the median spectral index  $\alpha=-0.29$ (in the convention $f_\nu \propto \nu^\alpha$), following \cite{gloudemans2021}. We obtain a radio-loudness of $R_{2500}=9\pm 1$, which places \quasarshort\ at the boundary between sources classified as radio-quiet or radio-loud (\citealt{kellermann1989,jiang2007}; but see also \citealt{calistro2024}). The uncertainty reported does not consider the uncertainty on the radio extrapolation. The rest-frame 144\,MHz specific radio luminosity is $L_{\rm 144}=(4.0 \pm 0.4) \times 10^{25}\,$W\,Hz$^{-1}$, similar to the bulk of $z>5$ quasars detected with LOFAR \citep[see e.g., Fig.~4 in][]{gloudemans2021}.  
If we assume a radio spectral index $\alpha=-0.7$ instead, the radio-loudness and 144\,MHz specific radio luminosity would be  $R_{2500}=4.6\pm 0.5$ and 
$L_{\rm 144}=(8.6 \pm 0.8) \times 10^{25}\,$W\,Hz$^{-1}$, respectively. 

We can conclude that \quasarshort\ has an intermediate radio-loudness of 
\mbox{$R_{2500}=4$--10} (depending on the radio spectral index; see above). However, the radio emission might not only come from synchrotron emission from the relativistic jet, because the quasar also shows evidence of outflows through high excitation BAL features (Sect.~\ref{sec:bal}): As shown by \cite{petley2022}, BAL quasars are more likely to be detected at 144\ MHz than their non-BAL counterparts, which suggests that shocks may cause part of the radio emission due to the BAL outflows interacting with the interstellar medium in their host galaxies.  Additional radio detections at other frequencies are required to interpret the radio properties of this source.

\section{Summary and outlook}\label{sec:outlook}

We have demonstrated that \textit{Euclid} slitless infrared spectroscopy is a powerful tool to identify quasars and to eliminate confusion with ultracool dwarfs by cross-correlating NISP \RGE\ spectra with templates.  The NISP \RGE\ spectral coverage is particularly well-matched to strong spectral features in L and T dwarfs (Fig.\ \ref{fig:a2769}, and \citealt{Q1-SP042}), but it can also help with the classifications for objects with less prominent spectral features in the 1206--$1892\,$nm spectral range, such as late M dwarfs (Fig.~\ref{fig:mdwarf}). Without the spectroscopic information, the ultracool dwarfs discussed here could mistakenly have been selected as high-redshift quasar candidates. Similar but more distant (thus fainter) brown dwarfs could incorrectly have been selected as high-redshift galaxies \citep[e.g.,][]{roberts-borsani2024}. \cite{EROLensData} argue that requiring $\IE - \YE>3$ reduces contamination by brown dwarfs. 
However, as shown in Fig.~\ref{fig:color-color}, late L- and T-dwarfs with such a significant colour break do exist and, therefore, brown dwarfs can still be a substantial contaminant to the $z>6$ galaxy candidates presented in \cite{EROLensVISDropouts}. 

In this paper, we focus on the highest redshift quasars to explore how efficiently \Euclid\ can help to fill the quasar redshift gaps at $z\approx 5.5$ and $z\gtrsim 7$, where NISP \RGE\ spectra allow us to identify two emission lines (Fig.~\ref{fig:motivation}).  
In the future, we will combine the grism data with photometric information (see Fig.~\ref{fig:color-color}). In that case, having even only one (or zero) strong emission line in the NISP \RGE\ spectra will help constrain the source redshift. 
Additionally, in the EDS, the expanded wavelength coverage provided by the NISP \BGE\ spectra can be effectively utilised for the reliable identification of sources (see Fig.~\ref{fig:motivation} and Appendix~\ref{sec:bluegrism}).

We have identified two sources in the EDF-N for which the best-fitting template is the quasar composite spectrum from \cite{vandenberk2001} at a redshift where two emission lines were expected. This blind experiment already showcases the potential of \Euclid\ for new quasar discoveries. The most promising source, with clear detections of the \ciii\ and \mgii\ lines, was confirmed as a quasar at the expected $z=5.4$ redshift with an optical follow-up spectrum (Fig.~\ref{fig:qso}). 
However, for the second candidate, no obvious emission lines were detected in the \Euclid spectrum or its follow-up, ground-based optical spectrum. The lack of a strong Lyman break rules out a source at $z\gtrsim 3.5$. 

The confirmation of \quasarshort\ at $z=5.4$ is noteworthy, especially considering that we anticipated finding fewer than one quasar of this kind in the surveyed area
\citep{schindler2023}. It is also important to point out that while Data Release 1 of the Dark Energy Spectroscopic Instrument (DESI) covered the EDF-N region and discovered hundreds of new quasars \citep{abdul-karim2025}, 
they excluded follow-up of $z\sim 5.4-5.6$ candidates due to a high expected contamination rate from M-dwarfs \citep{yang2023_desi}.
This emphasises the potential of NISP \RGE\ to effectively identify rare sources that may not be easily recognised through photometric methods. 
Across the entire EWS, this technique could potentially reveal around 400 quasars at $z\approx5.5$, similar to \quasarshort, as well as about 30 quasars at $z\gtrsim 7$.

A future improvement to a \Euclid\ grism-based selection is to include quasar templates with different spectral properties from those used in this work \cite[e.g.,][]{temple2021,EP-Lusso}. This can result in discovering additional quasars with different dust reddening or weaker/stronger emission lines than those of the average \citeauthor{vandenberk2001} quasar.

As a final note, \Euclid\ grism slitless spectroscopy will allow black hole mass measurements for thousands of (known and new) quasars, for which key emission lines such as \civ, \mgii, \hbeta, and \halpha, are not visible from the ground (Fig.~\ref{fig:motivation}).


\section*{Acknowledgements}
We would like to express our gratitude to the Euclid Consortium internal referees, Paul Hewett and Nicolas Lodieu, as well as MNRAS referee Matthew A.\ Malkan, for their constructive reviews, which significantly improved the quality of this paper. 
This work uses LBT data from programmes: LBT-2019A-I0211-0 and MPIA-2024A-003. 
We are grateful to Roland Gredel and the LBT staff for their support with the observations. 
ELM is supported by the European Research Council Advanced grant SUBSTELLAR, project number 101054354. 
This research has benefitted from the SpeX Prism Spectral Libraries, maintained by Adam Burgasser at https://cass.ucsd.edu/~ajb/browndwarfs/spexprism.
This work has made use of the Early Release
  Observations (ERO) data from the \Euclid\/ mission of the European
  Space Agency (ESA), 2024,
  \url{https://doi.org/10.57780/esa-qmocze3}.
We use the ERO dataset provided by ESA \citep{EROcite}. 
\AckEC
LOFAR data products were provided by the LOFAR Surveys Key Science project (LSKSP; https://lofar-surveys.org/) and were derived from observations with the International LOFAR Telescope (ILT). LOFAR \citep{vanHaarlem2013} is the Low Frequency Array designed and constructed by ASTRON. It has observing, data processing, and data storage facilities in several countries, which are owned by various parties (each with their own funding sources), and which are collectively operated by the ILT foundation under a joint scientific policy. The efforts of the LSKSP have benefited from funding from the European Research Council, NOVA, NWO, CNRS-INSU, the SURF Co-operative, the UK Science and Technology Funding Council and the Jülich Supercomputing Centre.
The Pan-STARRS1 Surveys (PS1) and the PS1 public science archive have been made possible through contributions by the Institute for Astronomy, the University of Hawaii, the Pan-STARRS Project Office, the Max-Planck Society and its participating institutes, the Max Planck Institute for Astronomy, Heidelberg and the Max Planck Institute for Extraterrestrial Physics, Garching, The Johns Hopkins University, Durham University, the University of Edinburgh, the Queen's University Belfast, the Harvard-Smithsonian Center for Astrophysics, the Las Cumbres Observatory Global Telescope Network Incorporated, the National Central University of Taiwan, the Space Telescope Science Institute, the National Aeronautics and Space Administration under Grant No. NNX08AR22G issued through the Planetary Science Division of the NASA Science Mission Directorate, the National Science Foundation Grant No. AST–1238877, the University of Maryland, Eotvos Lorand University (ELTE), the Los Alamos National Laboratory, and the Gordon and Betty Moore Foundation.
This project used public archival data from the Dark Energy Survey (DES). Funding for the DES Projects has been provided by the U.S. Department of Energy, the U.S. National Science Foundation, the Ministry of Science and Education of Spain, the Science and Technology FacilitiesCouncil of the United Kingdom, the Higher Education Funding Council for England, the National Center for Supercomputing Applications at the University of Illinois at Urbana-Champaign, the Kavli Institute of Cosmological Physics at the University of Chicago, the Center for Cosmology and Astro-Particle Physics at the Ohio State University, the Mitchell Institute for Fundamental Physics and Astronomy at Texas A\&M University, Financiadora de Estudos e Projetos, Funda{\c c}{\~a}o Carlos Chagas Filho de Amparo {\`a} Pesquisa do Estado do Rio de Janeiro, Conselho Nacional de Desenvolvimento Cient{\'i}fico e Tecnol{\'o}gico and the Minist{\'e}rio da Ci{\^e}ncia, Tecnologia e Inova{\c c}{\~a}o, the Deutsche Forschungsgemeinschaft, and the Collaborating Institutions in the Dark Energy Survey.
The Collaborating Institutions are Argonne National Laboratory, the University of California at Santa Cruz, the University of Cambridge, Centro de Investigaciones Energ{\'e}ticas, Medioambientales y Tecnol{\'o}gicas-Madrid, the University of Chicago, University College London, the DES-Brazil Consortium, the University of Edinburgh, the Eidgen{\"o}ssische Technische Hochschule (ETH) Z{\"u}rich,  Fermi National Accelerator Laboratory, the University of Illinois at Urbana-Champaign, the Institut de Ci{\`e}ncies de l'Espai (IEEC/CSIC), the Institut de F{\'i}sica d'Altes Energies, Lawrence Berkeley National Laboratory, the Ludwig-Maximilians Universit{\"a}t M{\"u}nchen and the associated Excellence Cluster Universe, the University of Michigan, the National Optical Astronomy Observatory, the University of Nottingham, The Ohio State University, the OzDES Membership Consortium, the University of Pennsylvania, the University of Portsmouth, SLAC National Accelerator Laboratory, Stanford University, the University of Sussex, and Texas A\&M University.
Based in part on observations at Cerro Tololo Inter-American Observatory, National Optical Astronomy Observatory, which is operated by the Association of Universities for Research in Astronomy (AURA) under a cooperative agreement with the National Science Foundation.
The Legacy Surveys consist of three individual and complementary projects: the Dark Energy Camera Legacy Survey (DECaLS; Proposal ID \#2014B-0404; PIs: David Schlegel and Arjun Dey), the Beijing-Arizona Sky Survey (BASS; NOAO Prop. ID \#2015A-0801; PIs: Zhou Xu and Xiaohui Fan), and the Mayall z-band Legacy Survey (MzLS; Prop. ID \#2016A-0453; PI: Arjun Dey). DECaLS, BASS and MzLS together include data obtained, respectively, at the Blanco telescope, Cerro Tololo Inter-American Observatory, NSF’s NOIRLab; the Bok telescope, Steward Observatory, University of Arizona; and the Mayall telescope, Kitt Peak National Observatory, NOIRLab. Pipeline processing and analyses of the data were supported by NOIRLab and the Lawrence Berkeley National Laboratory (LBNL). The Legacy Surveys project is honored to be permitted to conduct astronomical research on Iolkam Du’ag (Kitt Peak), a mountain with particular significance to the Tohono O’odham Nation.

NOIRLab is operated by the Association of Universities for Research in Astronomy (AURA) under a cooperative agreement with the National Science Foundation. LBNL is managed by the Regents of the University of California under contract to the U.S. Department of Energy.

This project used data obtained with the Dark Energy Camera (DECam), which was constructed by the Dark Energy Survey (DES) collaboration. Funding for the DES Projects has been provided by the U.S. Department of Energy, the U.S. National Science Foundation, the Ministry of Science and Education of Spain, the Science and Technology Facilities Council of the United Kingdom, the Higher Education Funding Council for England, the National Center for Supercomputing Applications at the University of Illinois at Urbana-Champaign, the Kavli Institute of Cosmological Physics at the University of Chicago, Center for Cosmology and Astro-Particle Physics at the Ohio State University, the Mitchell Institute for Fundamental Physics and Astronomy at Texas A\&M University, Financiadora de Estudos e Projetos, Fundacao Carlos Chagas Filho de Amparo, Financiadora de Estudos e Projetos, Fundacao Carlos Chagas Filho de Amparo a Pesquisa do Estado do Rio de Janeiro, Conselho Nacional de Desenvolvimento Cientifico e Tecnologico and the Ministerio da Ciencia, Tecnologia e Inovacao, the Deutsche Forschungsgemeinschaft and the Collaborating Institutions in the Dark Energy Survey. The Collaborating Institutions are Argonne National Laboratory, the University of California at Santa Cruz, the University of Cambridge, Centro de Investigaciones Energeticas, Medioambientales y Tecnologicas-Madrid, the University of Chicago, University College London, the DES-Brazil Consortium, the University of Edinburgh, the Eidgenossische Technische Hochschule (ETH) Zurich, Fermi National Accelerator Laboratory, the University of Illinois at Urbana-Champaign, the Institut de Ciencies de l’Espai (IEEC/CSIC), the Institut de Fisica d’Altes Energies, Lawrence Berkeley National Laboratory, the Ludwig Maximilians Universitat Munchen and the associated Excellence Cluster Universe, the University of Michigan, NSF’s NOIRLab, the University of Nottingham, the Ohio State University, the University of Pennsylvania, the University of Portsmouth, SLAC National Accelerator Laboratory, Stanford University, the University of Sussex, and Texas A\&M University.

BASS is a key project of the Telescope Access Program (TAP), which has been funded by the National Astronomical Observatories of China, the Chinese Academy of Sciences (the Strategic Priority Research Program “The Emergence of Cosmological Structures” Grant \# XDB09000000), and the Special Fund for Astronomy from the Ministry of Finance. The BASS is also supported by the External Cooperation Program of Chinese Academy of Sciences (Grant \# 114A11KYSB20160057), and Chinese National Natural Science Foundation (Grant \# 12120101003, \# 11433005).

The Legacy Survey team makes use of data products from the Near-Earth Object Wide-field Infrared Survey Explorer (NEOWISE), which is a project of the Jet Propulsion Laboratory/California Institute of Technology. NEOWISE is funded by the National Aeronautics and Space Administration.

The Legacy Surveys imaging of the DESI footprint is supported by the Director, Office of Science, Office of High Energy Physics of the U.S. Department of Energy under Contract No. DE-AC02-05CH1123, by the National Energy Research Scientific Computing Center, a DOE Office of Science User Facility under the same contract; and by the U.S. National Science Foundation, Division of Astronomical Sciences under Contract No. AST-0950945 to NOAO.

\section*{Data Availability}

 The data used in this study can be accessed from the observatories’ public archives or the websites of the surveys mentioned in the acknowledgments. The datasets generated and analysed during this study are available from the corresponding author upon reasonable request.

\section*{Affiliations}
\footnotesize{
$^{1}$Max-Planck-Institut f\"ur Astronomie, K\"onigstuhl 17, 69117 Heidelberg, Germany\\
$^{2}$Aix-Marseille Universit\'e, CNRS, CNES, LAM, Marseille, France\\
$^{3}$INAF-Osservatorio di Astrofisica e Scienza dello Spazio di Bologna, Via Piero Gobetti 93/3, 40129 Bologna, Italy\\
$^{4}$Jet Propulsion Laboratory, California Institute of Technology, 4800 Oak Grove Drive, Pasadena, CA, 91109, USA\\
$^{5}$Astrophysics Group, Blackett Laboratory, Imperial College London, London SW7 2AZ, UK\\
$^{6}$Department of Mathematics, Imperial College London, London SW7 2AZ, UK\\
$^{7}$Instituto de Astrof\'isica e Ci\^encias do Espa\c{c}o, Universidade do Porto, CAUP, Rua das Estrelas, PT4150-762 Porto, Portugal\\
$^{8}$DTx -- Digital Transformation CoLAB, Building 1, Azur\'em Campus, University of Minho, 4800-058 Guimar\~aes, Portugal\\
$^{9}$INAF-Osservatorio Astrofisico di Torino, Via Osservatorio 20, 10025 Pino Torinese (TO), Italy\\
$^{10}$Department of Physics, Astronomy and Mathematics, University of Hertfordshire, College Lane, Hatfield AL10 9AB, UK\\
$^{11}$School of Physics and Astronomy, University of Leicester, University Road, Leicester, LE1 7RH, UK\\
$^{12}$Departamento F\'isica Aplicada, Universidad Polit\'ecnica de Cartagena, Campus Muralla del Mar, 30202 Cartagena, Murcia, Spain\\
$^{13}$International Space University, 1 rue Jean-Dominique Cassini, 67400 Illkirch-Graffenstaden, France\\
$^{14}$Universit\'e de Strasbourg, CNRS, Observatoire astronomique de Strasbourg, UMR 7550, 67000 Strasbourg, France\\
$^{15}$Instituto de Astrof\'{\i}sica de Canarias, V\'{\i}a L\'actea, 38205 La Laguna, Tenerife, Spain\\
$^{16}$Universidad de La Laguna, Departamento de Astrof\'{\i}sica, 38206 La Laguna, Tenerife, Spain\\
$^{17}$Universite Marie et Louis Pasteur, CNRS, Observatoire des Sciences de l'Univers THETA Franche-Comte Bourgogne, Institut UTINAM, Observatoire de Besan\c con, BP 1615, 25010 Besan\c con Cedex, France\\
$^{18}$Universit\'e Claude Bernard Lyon 1, CNRS/IN2P3, IP2I Lyon, UMR 5822, Villeurbanne, F-69100, France\\
$^{19}$INAF-Osservatorio Astrofisico di Arcetri, Largo E. Fermi 5, 50125, Firenze, Italy\\
$^{20}$Dipartimento di Fisica e Astronomia, Universit\`{a} di Firenze, via G. Sansone 1, 50019 Sesto Fiorentino, Firenze, Italy\\
$^{21}$Research Center for Space and Cosmic Evolution, Ehime University, 2-5 Bunkyo-cho, Matsuyama, Ehime 790-8577, Japan\\
$^{22}$Department of Mathematics and Physics, Roma Tre University, Via della Vasca Navale 84, 00146 Rome, Italy\\
$^{23}$INAF-Osservatorio Astronomico di Roma, Via Frascati 33, 00078 Monteporzio Catone, Italy\\
$^{24}$Leiden Observatory, Leiden University, Einsteinweg 55, 2333 CC Leiden, The Netherlands\\
$^{25}$Hamburger Sternwarte, University of Hamburg, Gojenbergsweg 112, 21029 Hamburg, Germany\\
$^{26}$Department of Astronomy, University of Massachusetts, Amherst, MA 01003, USA\\
$^{27}$Max Planck Institute for Extraterrestrial Physics, Giessenbachstr. 1, 85748 Garching, Germany\\
$^{28}$Universit\"ats-Sternwarte M\"unchen, Fakult\"at f\"ur Physik, Ludwig-Maximilians-Universit\"at M\"unchen, Scheinerstrasse 1, 81679 M\"unchen, Germany\\
$^{29}$Dipartimento di Fisica e Astronomia "Augusto Righi" - Alma Mater Studiorum Universit\`a di Bologna, via Piero Gobetti 93/2, 40129 Bologna, Italy\\
$^{30}$INAF-IASF Milano, Via Alfonso Corti 12, 20133 Milano, Italy\\
$^{31}$INAF-Osservatorio Astronomico di Padova, Via dell'Osservatorio 5, 35122 Padova, Italy\\
$^{32}$INFN-Padova, Via Marzolo 8, 35131 Padova, Italy\\
$^{33}$Aix-Marseille Universit\'e, CNRS/IN2P3, CPPM, Marseille, France\\
$^{34}$Dipartimento di Fisica e Astronomia "G. Galilei", Universit\`a di Padova, Via Marzolo 8, 35131 Padova, Italy\\
$^{35}$Universit\'e Paris-Saclay, Universit\'e Paris Cit\'e, CEA, CNRS, AIM, 91191, Gif-sur-Yvette, France\\
$^{36}$Technical University of Munich, TUM School of Natural Sciences, Physics Department, James-Franck-Str.~1, 85748 Garching, Germany\\
$^{37}$Max-Planck-Institut f\"ur Astrophysik, Karl-Schwarzschild-Str.~1, 85748 Garching, Germany\\
$^{38}$Institut d'Astrophysique de Paris, UMR 7095, CNRS, and Sorbonne Universit\'e, 98 bis boulevard Arago, 75014 Paris, France\\
$^{39}$Centro de Astrobiolog\'ia (CAB), CSIC-INTA, ESAC Campus, Camino Bajo del Castillo s/n, 28692 Villanueva de la Ca\~nada, Madrid, Spain\\
$^{40}$Jodrell Bank Centre for Astrophysics, Department of Physics and Astronomy, University of Manchester, Oxford Road, Manchester M13 9PL, UK\\
$^{41}$Instituto de Astrof\'isica de Canarias (IAC); Departamento de Astrof\'isica, Universidad de La Laguna (ULL), 38200, La Laguna, Tenerife, Spain\\
$^{42}$Universit\'e Paris-Saclay, CNRS, Institut d'astrophysique spatiale, 91405, Orsay, France\\
$^{43}$Institute for Astronomy, University of Edinburgh, Royal Observatory, Blackford Hill, Edinburgh EH9 3HJ, UK\\
$^{44}$School of Physics, HH Wills Physics Laboratory, University of Bristol, Tyndall Avenue, Bristol, BS8 1TL, UK\\
$^{45}$Kavli Institute for the Physics and Mathematics of the Universe (WPI), University of Tokyo, Kashiwa, Chiba 277-8583, Japan\\
$^{46}$Center for Data-Driven Discovery, Kavli IPMU (WPI), UTIAS, The University of Tokyo, Kashiwa, Chiba 277-8583, Japan\\
$^{47}$School of Physics \& Astronomy, University of Southampton, Highfield Campus, Southampton SO17 1BJ, UK\\
$^{48}$Institute of Space Sciences (ICE, CSIC), Campus UAB, Carrer de Can Magrans, s/n, 08193 Barcelona, Spain\\
$^{49}$INAF-Istituto di Astrofisica e Planetologia Spaziali, via del Fosso del Cavaliere, 100, 00100 Roma, Italy\\
$^{50}$ESAC/ESA, Camino Bajo del Castillo, s/n., Urb. Villafranca del Castillo, 28692 Villanueva de la Ca\~nada, Madrid, Spain\\
$^{51}$School of Mathematics and Physics, University of Surrey, Guildford, Surrey, GU2 7XH, UK\\
$^{52}$INAF-Osservatorio Astronomico di Brera, Via Brera 28, 20122 Milano, Italy\\
$^{53}$IFPU, Institute for Fundamental Physics of the Universe, via Beirut 2, 34151 Trieste, Italy\\
$^{54}$INAF-Osservatorio Astronomico di Trieste, Via G. B. Tiepolo 11, 34143 Trieste, Italy\\
$^{55}$INFN, Sezione di Trieste, Via Valerio 2, 34127 Trieste TS, Italy\\
$^{56}$SISSA, International School for Advanced Studies, Via Bonomea 265, 34136 Trieste TS, Italy\\
$^{57}$Dipartimento di Fisica e Astronomia, Universit\`a di Bologna, Via Gobetti 93/2, 40129 Bologna, Italy\\
$^{58}$INFN-Sezione di Bologna, Viale Berti Pichat 6/2, 40127 Bologna, Italy\\
$^{59}$Centre National d'Etudes Spatiales -- Centre spatial de Toulouse, 18 avenue Edouard Belin, 31401 Toulouse Cedex 9, France\\
$^{60}$Space Science Data Center, Italian Space Agency, via del Politecnico snc, 00133 Roma, Italy\\
$^{61}$Dipartimento di Fisica, Universit\`a di Genova, Via Dodecaneso 33, 16146, Genova, Italy\\
$^{62}$INFN-Sezione di Genova, Via Dodecaneso 33, 16146, Genova, Italy\\
$^{63}$Department of Physics "E. Pancini", University Federico II, Via Cinthia 6, 80126, Napoli, Italy\\
$^{64}$INAF-Osservatorio Astronomico di Capodimonte, Via Moiariello 16, 80131 Napoli, Italy\\
$^{65}$Faculdade de Ci\^encias da Universidade do Porto, Rua do Campo de Alegre, 4150-007 Porto, Portugal\\
$^{66}$Dipartimento di Fisica, Universit\`a degli Studi di Torino, Via P. Giuria 1, 10125 Torino, Italy\\
$^{67}$INFN-Sezione di Torino, Via P. Giuria 1, 10125 Torino, Italy\\
$^{68}$Centro de Investigaciones Energ\'eticas, Medioambientales y Tecnol\'ogicas (CIEMAT), Avenida Complutense 40, 28040 Madrid, Spain\\
$^{69}$Port d'Informaci\'{o} Cient\'{i}fica, Campus UAB, C. Albareda s/n, 08193 Bellaterra (Barcelona), Spain\\
$^{70}$Institute for Theoretical Particle Physics and Cosmology (TTK), RWTH Aachen University, 52056 Aachen, Germany\\
$^{71}$INFN section of Naples, Via Cinthia 6, 80126, Napoli, Italy\\
$^{72}$Dipartimento di Fisica e Astronomia "Augusto Righi" - Alma Mater Studiorum Universit\`a di Bologna, Viale Berti Pichat 6/2, 40127 Bologna, Italy\\
$^{73}$European Space Agency/ESRIN, Largo Galileo Galilei 1, 00044 Frascati, Roma, Italy\\
$^{74}$Institute of Physics, Laboratory of Astrophysics, Ecole Polytechnique F\'ed\'erale de Lausanne (EPFL), Observatoire de Sauverny, 1290 Versoix, Switzerland\\
$^{75}$Institut de Ci\`{e}ncies del Cosmos (ICCUB), Universitat de Barcelona (IEEC-UB), Mart\'{i} i Franqu\`{e}s 1, 08028 Barcelona, Spain\\
$^{76}$Instituci\'o Catalana de Recerca i Estudis Avan\c{c}ats (ICREA), Passeig de Llu\'{\i}s Companys 23, 08010 Barcelona, Spain\\
$^{77}$UCB Lyon 1, CNRS/IN2P3, IUF, IP2I Lyon, 4 rue Enrico Fermi, 69622 Villeurbanne, France\\
$^{78}$Mullard Space Science Laboratory, University College London, Holmbury St Mary, Dorking, Surrey RH5 6NT, UK\\
$^{79}$Canada-France-Hawaii Telescope, 65-1238 Mamalahoa Hwy, Kamuela, HI 96743, USA\\
$^{80}$Departamento de F\'isica, Faculdade de Ci\^encias, Universidade de Lisboa, Edif\'icio C8, Campo Grande, PT1749-016 Lisboa, Portugal\\
$^{81}$Instituto de Astrof\'isica e Ci\^encias do Espa\c{c}o, Faculdade de Ci\^encias, Universidade de Lisboa, Campo Grande, 1749-016 Lisboa, Portugal\\
$^{82}$Department of Astronomy, University of Geneva, ch. d'Ecogia 16, 1290 Versoix, Switzerland\\
$^{83}$Dipartimento di Fisica "Aldo Pontremoli", Universit\`a degli Studi di Milano, Via Celoria 16, 20133 Milano, Italy\\
$^{84}$INFN-Sezione di Milano, Via Celoria 16, 20133 Milano, Italy\\
$^{85}$Institute of Theoretical Astrophysics, University of Oslo, P.O. Box 1029 Blindern, 0315 Oslo, Norway\\
$^{86}$Department of Physics, Lancaster University, Lancaster, LA1 4YB, UK\\
$^{87}$Felix Hormuth Engineering, Goethestr. 17, 69181 Leimen, Germany\\
$^{88}$Technical University of Denmark, Elektrovej 327, 2800 Kgs. Lyngby, Denmark\\
$^{89}$Cosmic Dawn Center (DAWN), Denmark\\
$^{90}$NASA Goddard Space Flight Center, Greenbelt, MD 20771, USA\\
$^{91}$Department of Physics and Astronomy, University College London, Gower Street, London WC1E 6BT, UK\\
$^{92}$Department of Physics and Helsinki Institute of Physics, Gustaf H\"allstr\"omin katu 2, 00014 University of Helsinki, Finland\\
$^{93}$Universit\'e de Gen\`eve, D\'epartement de Physique Th\'eorique and Centre for Astroparticle Physics, 24 quai Ernest-Ansermet, CH-1211 Gen\`eve 4, Switzerland\\
$^{94}$Department of Physics, P.O. Box 64, 00014 University of Helsinki, Finland\\
$^{95}$Helsinki Institute of Physics, Gustaf H{\"a}llstr{\"o}min katu 2, University of Helsinki, Helsinki, Finland\\
$^{96}$NOVA optical infrared instrumentation group at ASTRON, Oude Hoogeveensedijk 4, 7991PD, Dwingeloo, The Netherlands\\
$^{97}$Centre de Calcul de l'IN2P3/CNRS, 21 avenue Pierre de Coubertin 69627 Villeurbanne Cedex, France\\
$^{98}$Universit\"at Bonn, Argelander-Institut f\"ur Astronomie, Auf dem H\"ugel 71, 53121 Bonn, Germany\\
$^{99}$INFN-Sezione di Roma, Piazzale Aldo Moro, 2 - c/o Dipartimento di Fisica, Edificio G. Marconi, 00185 Roma, Italy\\
$^{100}$Department of Physics, Institute for Computational Cosmology, Durham University, South Road, Durham, DH1 3LE, UK\\
$^{101}$Universit\'e Paris Cit\'e, CNRS, Astroparticule et Cosmologie, 75013 Paris, France\\
$^{102}$CNRS-UCB International Research Laboratory, Centre Pierre Bin\'etruy, IRL2007, CPB-IN2P3, Berkeley, USA\\
$^{103}$Institut d'Astrophysique de Paris, 98bis Boulevard Arago, 75014, Paris, France\\
$^{104}$Aurora Technology for European Space Agency (ESA), Camino bajo del Castillo, s/n, Urbanizacion Villafranca del Castillo, Villanueva de la Ca\~nada, 28692 Madrid, Spain\\
$^{105}$Institut de F\'{i}sica d'Altes Energies (IFAE), The Barcelona Institute of Science and Technology, Campus UAB, 08193 Bellaterra (Barcelona), Spain\\
$^{106}$European Space Agency/ESTEC, Keplerlaan 1, 2201 AZ Noordwijk, The Netherlands\\
$^{107}$School of Mathematics, Statistics and Physics, Newcastle University, Herschel Building, Newcastle-upon-Tyne, NE1 7RU, UK\\
$^{108}$DARK, Niels Bohr Institute, University of Copenhagen, Jagtvej 155, 2200 Copenhagen, Denmark\\
$^{109}$Waterloo Centre for Astrophysics, University of Waterloo, Waterloo, Ontario N2L 3G1, Canada\\
$^{110}$Department of Physics and Astronomy, University of Waterloo, Waterloo, Ontario N2L 3G1, Canada\\
$^{111}$Perimeter Institute for Theoretical Physics, Waterloo, Ontario N2L 2Y5, Canada\\
$^{112}$Institute of Space Science, Str. Atomistilor, nr. 409 M\u{a}gurele, Ilfov, 077125, Romania\\
$^{113}$Consejo Superior de Investigaciones Cientificas, Calle Serrano 117, 28006 Madrid, Spain\\
$^{114}$Institut f\"ur Theoretische Physik, University of Heidelberg, Philosophenweg 16, 69120 Heidelberg, Germany\\
$^{115}$Institut de Recherche en Astrophysique et Plan\'etologie (IRAP), Universit\'e de Toulouse, CNRS, UPS, CNES, 14 Av. Edouard Belin, 31400 Toulouse, France\\
$^{116}$Universit\'e St Joseph; Faculty of Sciences, Beirut, Lebanon\\
$^{117}$Departamento de F\'isica, FCFM, Universidad de Chile, Blanco Encalada 2008, Santiago, Chile\\
$^{118}$Universit\"at Innsbruck, Institut f\"ur Astro- und Teilchenphysik, Technikerstr. 25/8, 6020 Innsbruck, Austria\\
$^{119}$Institut d'Estudis Espacials de Catalunya (IEEC),  Edifici RDIT, Campus UPC, 08860 Castelldefels, Barcelona, Spain\\
$^{120}$Satlantis, University Science Park, Sede Bld 48940, Leioa-Bilbao, Spain\\
$^{121}$Infrared Processing and Analysis Center, California Institute of Technology, Pasadena, CA 91125, USA\\
$^{122}$Instituto de Astrof\'isica e Ci\^encias do Espa\c{c}o, Faculdade de Ci\^encias, Universidade de Lisboa, Tapada da Ajuda, 1349-018 Lisboa, Portugal\\
$^{123}$Cosmic Dawn Center (DAWN)\\
$^{124}$Niels Bohr Institute, University of Copenhagen, Jagtvej 128, 2200 Copenhagen, Denmark\\
$^{125}$Universidad Polit\'ecnica de Cartagena, Departamento de Electr\'onica y Tecnolog\'ia de Computadoras,  Plaza del Hospital 1, 30202 Cartagena, Spain\\
$^{126}$INFN-Bologna, Via Irnerio 46, 40126 Bologna, Italy\\
$^{127}$Kapteyn Astronomical Institute, University of Groningen, PO Box 800, 9700 AV Groningen, The Netherlands\\
$^{128}$INAF, Istituto di Radioastronomia, Via Piero Gobetti 101, 40129 Bologna, Italy\\
$^{129}$Department of Physics, Oxford University, Keble Road, Oxford OX1 3RH, UK\\
$^{130}$ICL, Junia, Universit\'e Catholique de Lille, LITL, 59000 Lille, France\\
$^{131}$ICSC - Centro Nazionale di Ricerca in High Performance Computing, Big Data e Quantum Computing, Via Magnanelli 2, Bologna, Italy\\
$^{132}$Department of Physics and Astronomy, University of British Columbia, Vancouver, BC V6T 1Z1, Canada
}}



\bibliographystyle{mnras}
\bibliography{Euclid,library_nameyyyy,Q1} 




\appendix

\section{Ultracool dwarf templates}
\label{apdx:A}
The spectral benchmarks used as templates in this work are listed in Table~\ref{tab:spectral_names}.

\begin{table}
\newcommand{\pd}{\phantom{1}}
\setlength{\tabcolsep}{3.25pt}
\centering
\caption{Spectral benchmarks used as templates in this work}
\begin{tabular}{|llr|}
\hline
\textbf{Spectral Type} & \textbf{Name} & \textbf{Reference} \\
  & &  \\[-8pt]
\hline
  & &  \\[-8pt]
M6 & LHS 36 & \cite{burgasser2008} \\
M7 & ITG2& \cite{muench2007} \\
M8 & KPNO6 & \cite{muench2007}  \\
M9 & KPNO12 & \cite{muench2007} \\
L0 & 2MASSJ12474944$-$1117551 & \cite{kirkpatrick2010} \\
L1 & 2MASSJ14313097+1436539 & \cite{sheppard2009} \\
L2 & 2MASSJ01415823-4633574& \cite{kirkpatrick2006} \\
L3 & SDSSJ213352.72+101841.0 & \cite{chiu2006} \\
L4 & 2MASSJ03001631+2130205 & \cite{kirkpatrick2010} \\
L5 & 2MASSIJ1526140+204341 & \cite{burgasser2004}\\
L6 & SDSSJ134203.11+134022.2 & \cite{chiu2006}\\
L7 & 2MASSJ21481628+4003593 & \cite{looper2008} \\
L8 & 2MASSJ10430758+2225236 & \cite{siegler2007} \\
L9 & SDSSJ213154.43$-$011939.3 & \cite{chiu2006} \\
T0 & Gl337CD& \cite{burgasser2010}\\
T1 & SDSSJ015141.69+124429.6 & \cite{burgasser2004} \\
T2 & 2MASSJ15461461+4932114 & \cite{burgasser2010}\\
T3 & SDSSJ153417.05+161546.1AB & \cite{chiu2006} \\
T4 & 2MASSJ10595219+3041498 & \cite{sheppard2009}\\
T5 & 2MASSJ18283572-4849046 & \cite{burgasser2004} \\
T6 & 2MASSJ16150413+1340079 & \cite{looper2007} \\
T7 & 2MASSJ00501994-3322402 & \cite{burgasser2006} \\
T8 & 2MASSJ09393548-2448279 & \cite{burgasser2006d} \\
\hline
\end{tabular}
\label{tab:spectral_names}
\end{table}

\section{\Euclid\ NISP \RGE\ 2D spectrograms}
\label{sec:2dspec}

\begin{figure*}
    \centering
    \includegraphics[width=\textwidth]{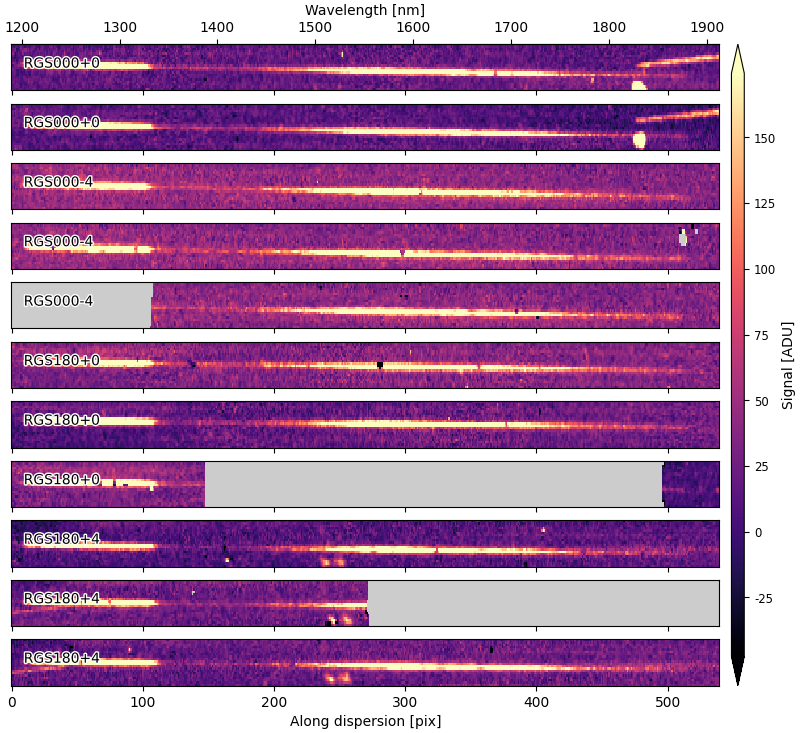}
    \caption{
    NISP two-dimensional spectrograms of \dwarfabell. The one-dimensional extraction is shown in Fig.~\ref{fig:a2769}. Note that this source is from the Abell 2766 ERO program and, therefore has three times more data than the sources from the EWS. Grey regions correspond to missing data.
    }
    \label{fig:twod_ero}
\end{figure*}

Since this is one of the first publications including NISP \RGE\ spectra from an ERO program \citep{EROcite, EROLensData}, we also provide in Fig.~\ref{fig:twod_ero} the two-dimensional spectrogram for the T dwarf discussed in Sect.~\ref{sec:a2764}. 
Note that these two-dimensional data are not standard products of the \Euclid\ pipeline and were processed with a custom pipeline as described in Sect.~\ref{sec:grism}.

\section{\Euclid\ NISP \BGE\ early data for $z\approx 5.5$ candidates in the EDF-N}
\label{sec:bluegrism}
As noted in Sect.~\ref{sec:grism}, the NISP \BGE\ data from the EDF-N was made available during the final stages of this manuscript. Here, we present the NISP \BGE\ for the sources discussed in  Sect.~\ref{sec:discovery-quasars}, showcasing one of the first scientific demonstrations of NISP \BGE\ data.  

\begin{figure*}
    \centering
    \includegraphics[width=\textwidth]{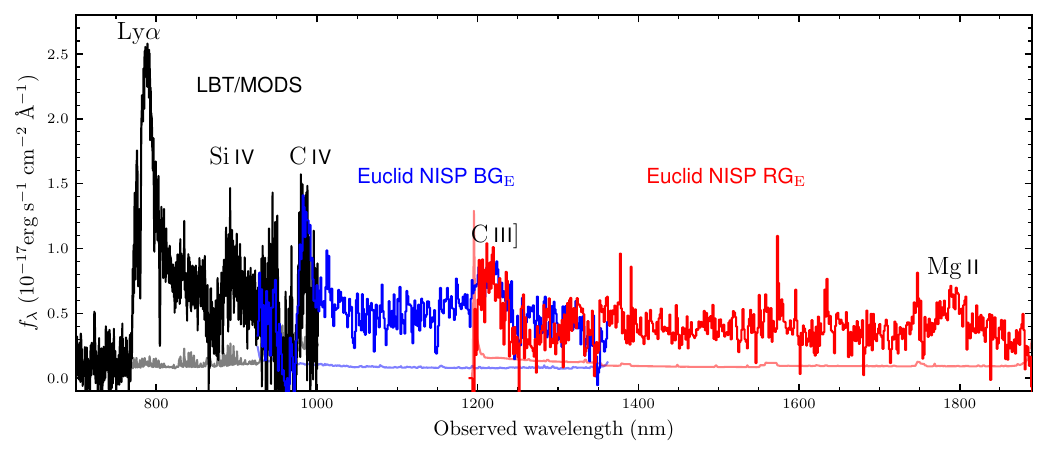}
    \caption{
    Spectra of the $z=5.4$ quasar \quasarshort. The LBT/MODS (black) and the NISP \RGE\ (red) spectra were shown in Fig.~\ref{fig:qso}. 
    The NISP \BGE\ spectrum covers the wavelength range that connects the other two spectra. The uncertainties in the spectra are represented in lighter colours corresponding to each spectrum.}
    \label{fig:qso-blue}
\end{figure*}

Figure~\ref{fig:qso-blue} shows all the available spectra for the quasar \quasarshort. The NISP \BGE\ bridges the LBT/MODS and NISP \RGE\ spectra shown in Fig.~\ref{fig:qso}, and the overlapping regions are consistent with each other. 
From both Euclid NISP spectra, three strong emission lines are identified at $z=5.4$: \civ, \ciii, and \mgii. The \civ\ BAL is evident in the NISP \BGE\ spectrum and consistent with the measurements from the LBT/MODS spectrum (Fig.~\ref{fig:bal}).

\begin{figure*}
    \centering
    \includegraphics[width=\textwidth]{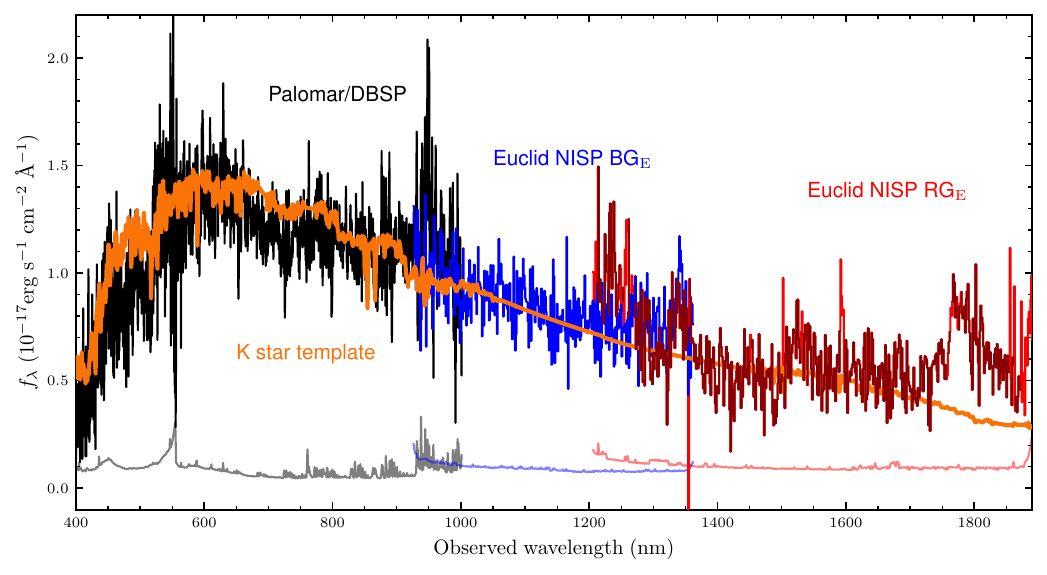}
    \caption{
 Spectra of the quasar candidate \noquasar\ (based solely on NISP \RGE\ data). The Palomar/DBSP (black) and the NISP \RGE\ (red) spectra were shown in Fig.~\ref{fig:qso-contaminant}. 
    The NISP \BGE\ spectrum covers the wavelength range that connects the other two spectra. The uncertainties in the spectra are represented in lighter colours corresponding to each spectrum. 
    The orange line represents the best template identified by the \Euclid\ pipeline, considering both NISP \BGE\ and \RGE\ spectra, corresponding to a K-type star.
    }
    \label{fig:cont-blue}
    \label{lastpage}
\end{figure*}

Figure~\ref{fig:cont-blue} shows all the available spectra for the source \noquasar, which was ruled out to be a $z\approx 5.5$ quasar in Sect.~\ref{sec:noquasar}. The NISP \BGE\ bridges the Palomar/DBSP and NISP \RGE\ spectra shown in Fig.~\ref{fig:qso-contaminant}, and the overlapping regions are consistent with each other.
When analysing both the NISP \BGE\ and \RGE\ data, the best template fit identified by the \Euclid pipeline is a K star, as depicted in orange in Fig.~\ref{fig:cont-blue}). The K star template closely matches the overall shape of the spectra, including the Palomar/DBSP spectrum, which was not utilised for classification. However, this template fails to account for the broad emission lines observed in the NISP \RGE\ spectrum at approximately 1200\,nm and 1800\,nm, which initially suggested a quasar classification. These prominent features, which are broader than the spectral resolution, are puzzling. Fortunately, this source is in the EDF-N, meaning that multiple epochs will be available to verify if the features in this spectrum are real.

\bsp	
\end{document}